\documentclass[aps,floatfix,superscriptaddress,preprintnumbers,showpacs,showkeys,nofootinbib]{revtex4}
\usepackage{bm}
\usepackage{graphicx}
\usepackage{amssymb}
\usepackage{amsmath}
\usepackage{color}
\usepackage{glossaries}
\usepackage{units}
\usepackage{subfigure}
\usepackage{slashed,braket}

\newcommand{\be}{\begin{equation}}
\newcommand{\ee}{\end{equation}}
\newcommand{\ba}{\begin{eqnarray}}
\newcommand{\ea}{\end{eqnarray}}
\newcommand{\nn}{\nonumber}
\newcommand{\sle}{\slashed \epsilon}
\newcommand{\slk}{\slashed k}
\newcommand{\slp}{\slashed p}
\newcommand{\slq}{\slashed q}
\newcommand{\slz}{\slashed z}
\newcommand{\sla}{\slashed}
\newcommand{\eq}{\begin{eqnarray}}
\newcommand{\en}{\end{eqnarray}}

\newacronym{chpt}{ChPT}{Chiral Perturbation Theory}
\newacronym[plural=EFTs,firstplural=effective field theories (EFTs)]{eft}{EFT}{effective field theory}
\newacronym{qcd}{QCD}{quantum chromodynamics}
\newacronym{hbchpt}{HBChPT}{heavy-baryon ChPT}
\newacronym{mami}{MAMI}{Mainz Microtron}
\newacronym[plural=LECs,firstplural=low-energy constants (LECs)]{lec}{LEC}{low-energy constant}
\newacronym[plural=LETs,firstplural=low-energy theorems (LETs)]{let}{LET}{low-energy theorem}
\newacronym{eoms}{EOMS}{Extended On-Mass Shell}
\newacronym{cglm}{CGLM}{Chew--Goldberger--Low--Nambu}
\newacronym{wfr}{WFR}{wave-function renormalization}
\newacronym{rcs}{RCS}{real Compton scattering}
\newacronym{vcs}{VCS}{virtual Compton scattering}
\newacronym{gdh}{GDH}{Gerasimov--Drell--Hearn}
\newacronym{sd}{SD}{spin dependent}
\newacronym{si}{SI}{spin independent}
\newacronym{ggt}{GGT}{Gell-Mann, Goldberger and Thirring}
\newacronym{edm}{EDM}{electric dipole moment}
\newacronym{cl}{CL}{confidence level}
\newacronym{fesr}{FESR}{finite-energy sum rules}
\newacronym{sse}{SSE}{small scale expansion}
\newacronym{sm}{SM}{Standard Model}
\newacronym{qed}{QED}{quantum electrodynamics} 
\newacronym{pcbt}{PCBT}{power-counting breaking terms}

\begin{document}

\title{$CP$-violating decays of the pseudoscalars $\eta$ and $\eta'$ \\ 
and their connection to the electric dipole moment of the neutron}

\author{Thomas~Gutsche}
\affiliation{Institut f\"ur Theoretische Physik,
Universit\"at T\"ubingen,
Kepler Center for Astro and Particle Physics,
Auf der Morgenstelle 14, D-72076 T\"ubingen, Germany}
\author{Astrid~N.~Hiller Blin} 
\affiliation{Instituto de F\'{\i}sica Corpuscular, Universidad de Valencia--CSIC,\\
Institutos de Investigaci\'on, Ap. Correos 22085, E-46071 Valencia, Spain} 
\author{Sergey~Kovalenko}
\affiliation{Departamento de F\'\i sica y Centro Cient\'\i fico
Tecnol\'ogico de Valpara\'\i so (CCTVal), Universidad T\'ecnica
Federico Santa Mar\'\i a, Casilla 110-V, Valpara\'\i so, Chile}
\author{Serguei~Kuleshov}
\affiliation{Departamento de F\'\i sica y Centro Cient\'\i fico
Tecnol\'ogico de Valpara\'\i so (CCTVal), Universidad T\'ecnica
Federico Santa Mar\'\i a, Casilla 110-V, Valpara\'\i so, Chile}
\author{Valery~E.~Lyubovitskij}
\affiliation{Institut f\"ur Theoretische Physik,
Universit\"at T\"ubingen,
Kepler Center for Astro and Particle Physics,
Auf der Morgenstelle 14, D-72076 T\"ubingen, Germany}
\affiliation{Department of Physics, Tomsk State University, 634050 Tomsk, Russia}
\affiliation{Laboratory of Particle Physics, Mathematical Physics Department,
Tomsk Polytechnic University, 634050 Tomsk, Russia}
\author{Manuel J.~Vicente Vacas}
\affiliation{Instituto de F\'{\i}sica Corpuscular, Universidad de Valencia--CSIC,\\
Institutos de Investigaci\'on, Ap. Correos 22085, E-46071 Valencia, Spain} 
\author{Alexey~Zhevlakov} 
\affiliation{Department of Physics, Tomsk State University, 634050 Tomsk, Russia}

\today

\begin{abstract}

Using the present upper bound on the neutron electric dipole moment, we give an estimate for the upper limit 
of the $CP$-violating couplings of the $\eta(\eta')$ to the nucleon. Using this result, we then 
derive constraints on the $CP$-violating $\eta(\eta')\pi\pi$ couplings, which 
define the two-pion $CP$-violating decays of the $\eta$ and $\eta'$ mesons. Our results 
are relevant for the running and planned measurements of rare decays of 
the $\eta$ and $\eta'$ mesons by the GlueX Collaboration at JLab and the 
LHCb Collaboration at CERN. 

\end{abstract}



\maketitle

\section{Introduction}
\label{SEDMIntro}

The $CP$ violation (CPV) is crucial for understanding the observed baryon asymmetry of the Universe (BAU).
In the Standard Model (SM), $CP$ is explicitly broken by the complex phase of the 
Cabibbo-Kobayashi-Maskawa (CKM) quark-mixing matrix and by the $\theta$-term of QCD. Up to date, experimentally 
CPV has only been observed in $K$- and $B$-meson mixing and hadronic decays~\cite{Agashe:2014kda},
which are perfectly compatible with the CKM phase. 
On the other hand, the CPV of SM origin is by far 
insufficient for the explanation of the BAU.  
The missing amount of CPV is believed to arise from non-SM sources. 

Apart from the above-mentioned CPV observables, there are others with 
distinct sensitivity to different sources of CPV. 
Among them, the electric dipole moments (EDMs) of the neutron, leptons and atoms 
have attracted special attention~\cite{Pich:1991fq,Dar:2000tn,Gorchtein:2008pe,Jarlskog:1995ww}.
In particular, the neutron EDM is weakly sensitive to the CKM phase, 
but strongly sensitive to the $\theta$-term, constraining the latter 
to be unnaturally small. This smallness is elegantly explained 
by the famous Peccei and Quinn mechanism~\cite{Peccei:1977hh,Peccei:1977ur}.

Various beyond-the-SM contributions to the EDM have been studied in the literature, for example,  the $R$-parity violating 
supersymmetry~\cite{Faessler:2006at,Faessler:2006vi} and 
meson-cloud effects in the nucleon~\cite{Kuckei:2005pg,Dib:2006hk}.  
For a review on EDMs as probes of new physics see, e.g., Ref.~\cite{Pospelov:2005pr}. 

There is also an extensive experimental program, both for the measurements of the EDMs, and 
looking for rare CPV decays with increasing sensitivity. 
In particular, searches for rare $\eta$ and $\eta'$ decays have been performed 
by the LHCb Collaboration at CERN~\cite{Aaij:2016jaa}, and are planned 
by the GlueX experiment at JLab (Hall D)~\cite{Ghoul:2015ifw}. 
In the present paper we focus on $\eta (\eta^{'}) \rightarrow\pi\pi$.  
As will be shown, the CPV $\eta(\eta')\pi\pi$-couplings underlying these decays
also contribute to the neutron EDM. Thus, the current experimental limits on 
the neutron EDM~\cite{Agashe:2014kda} 
\begin{eqnarray}\label{eq:Neutron-EDM-exp}
|d_{n}| \leq 2.9\times 10^{-26}e~\unit{cm},  \ \ \ 90\% \ {\rm C.L.},
\end{eqnarray}
will allow us to derive new  indirect upper bounds on the branching ratios of 
these CPV decays.  
The current direct experimental 90\% C.L. upper 
limits~\cite{Agashe:2014kda,Aaij:2016jaa} are
\begin{align}
\nn\frac{\Gamma(\eta\rightarrow\pi\pi)}{\Gamma_\eta^{\text{tot}}}
&<
\Bigg\{\begin{array}{c}
1.3\times 10^{-5}\text{ for }\pi^+\pi^-\\
3.5\times 10^{-4}\text{ for }\pi^0\pi^0
\end{array},\\
\frac{\Gamma(\eta'\rightarrow\pi\pi)}{\Gamma_{\eta'}^{\text{tot}}}
&<
\Bigg\{\begin{array}{c}
1.8\times 10^{-5}\text{ for }\pi^+\pi^-\\
4.0\times 10^{-4}\text{ for }\pi^0\pi^0
\end{array}\,.\label{Epdgbranch}
\end{align}
Here $\Gamma_\eta^{\text{tot}}=(1.31\pm0.05)\,\unit{keV}$ 
and $\Gamma_{\eta'}^{\text{tot}}=(0.198\pm0.009)\,\unit{MeV}$ 
are the total decay widths.

In Ref.~\cite{Pich:1991fq}, the size of the neutron \gls{edm} was estimated 
on the basis of a CPV chiral Lagrangian that couples the 
light pseudoscalars to the neutron, modulo the CPV phase. 
At leading order, only the contributions of the charged mesons survive, 
for which there is no experimental input on the size of their CPV couplings. 
In order to relate the neutron \gls{edm} to the couplings with the $\eta(\eta')$, 
next-to-leading order chiral Lagrangians must be taken into account. 

This is one of the aims of the present work. We carry out the analysis of the EDM within 
fully covariant \gls{chpt}, and with the explicit inclusion of intermediate 
spin-$3/2$ states, namely the $\Delta(1232)$ resonance. The latter couples strongly 
to the nucleon, and is therefore expected to give important contributions to processes 
that lie in energies close to the resonance mass. We use the \gls{eoms} scheme~\cite{Gegelia:1999gf,Fuchs:2003qc}
 for renormalization. 
It is relativistic, satisfies analyticity, and usually converges faster than non-relativistic 
approaches.

The paper is organized as follows. 
In Sec.~\ref{STheoCPetapipi}, we construct the Lagrangian for the $CP$-violating coupling 
of the $\eta(\eta')$ to the pions, in order to connect it 
with the branching ratio of the reaction. In Sec.~\ref{STheoCPetaNN}, we use this input to construct the 
$CP$-violating coupling of the $\eta(\eta')$ to the nucleon. The $CP$-conserving coupling 
is discussed in Sec.~\ref{STheoetaNN}, with the usual chiral Lagrangian considerations. 
In Sec.~\ref{STheoVec}, we give a brief overview on the couplings with vector mesons. 
The calculation of an estimate for the neutron \gls{edm} with these tools 
is shown in Sec.~\ref{STheoEDM}. By comparing the result with the experimental constraint 
on the \gls{edm}, we extract an estimate for the $\eta(\eta')\rightarrow\pi\pi$ branching 
ratio upper limits. 
Finally, in Sec.~\ref{Ssummedm}, we summarize the work and give our conclusions.

\section{The $CP$ violating $\eta(\eta')\rightarrow\pi\pi$ decay}\label{STheoCPetapipi}

The effective Lagrangian describing the $CP$-violating $\eta(\eta') \pi\pi$ coupling 
is given by ~\cite{Gorchtein:2008pe}:
\begin{align}
\mathcal{L}^\text{CP}_{H\pi\pi}=f_{H\pi\pi} \, 
M_H \, H \, \vec{\pi\,}^2\,, \label{ELagCPepp}
\end{align} 
with $H=\eta,\eta'$, $M_H$ the mass of the $\eta(\eta')$ meson and 
$f_{H\pi\pi}$ the coupling constant of $\eta(\eta')$ 
to the pions. Thus, the decay width is given by
\begin{align}
\Gamma=\frac{n_\Gamma \, 
|\vec{p}_\pi|}{8\pi \, m_H^2}\,\left|\mathcal{M}_{H\pi\pi}\right|^2
= n_\Gamma \, \frac{\sqrt{M_H^2-4M_\pi^2}}{4\pi} \, |f_{H\pi\pi}|^2\,, 
\end{align} 
where $n_\Gamma$ is an additional final-state factor, 
which equals 1/2 for the $\pi^0\pi^0$ and $1$ for the $\pi^+\pi^-$ channel. 
Using the limits from Eq.~\ref{Epdgbranch}, we obtain upper limits for the coupling constants.

Here we choose to calculate the charged and neutral channels separately, 
and to keep only the lower result as the global upper limit: 
\begin{align}
\nn |f_{\eta\pi\pi}|&<2.1\times10^{-5},\\
|f_{\eta'\pi\pi}|&<2.2\times10^{-4}\,.
\label{upperbound}
\end{align}

\section{The $CP$-violating couplings of the $\eta$ and $\eta'$ to the nucleon}
\label{STheoCPetaNN}

\begin{figure}[htbp]
\begin{center}
\subfigure[]{
\label{fetaNloopa}
\includegraphics[width=0.3\textwidth]{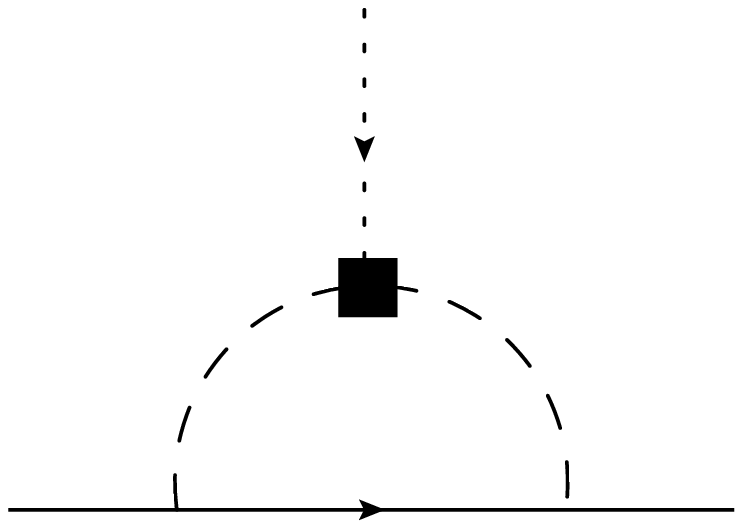}}
\subfigure[]{
\label{fetaNloopaD}
\includegraphics[width=0.3\textwidth]{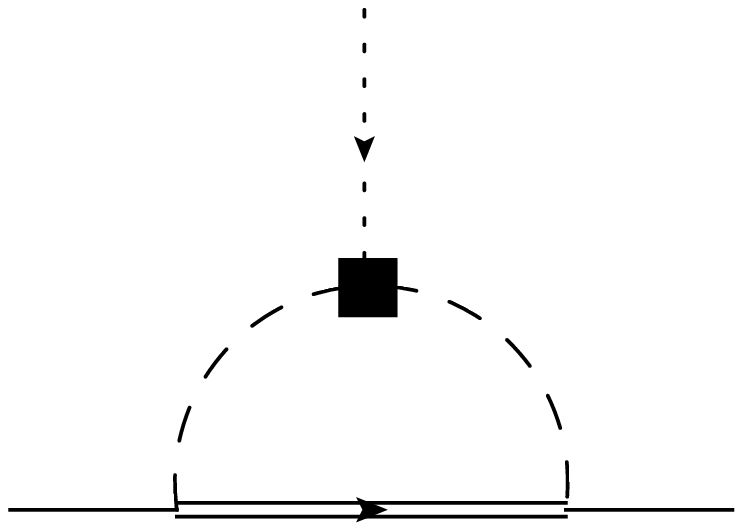}}
\subfigure[]{
\label{fetaNloopb}
\includegraphics[width=0.3\textwidth]{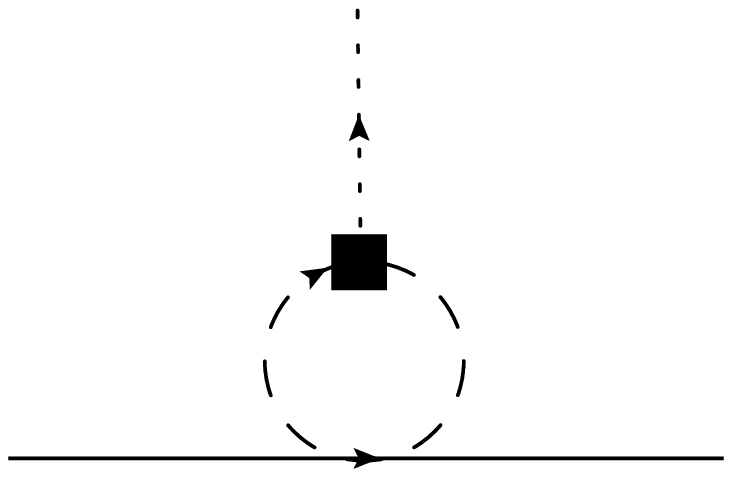}}
\end{center}
\caption{Loops that can contribute to the $CP$-violating coupling of the $\eta(\eta')$ 
to the nucleon. The single solid lines stand for nucleons, the double lines for the 
$\Delta$, the dashes for pions and the dotted lines for the $\eta(\eta')$. 
The black box at the $H\pi\pi$ vertex indicates the $CP$-violating coupling.}
\label{fetaNloop}
\end{figure}

With the previous considerations, one can obtain an estimate for 
the $CP$-violating coupling of the $\eta(\eta')$ to the nucleon
\begin{align}
\mathcal{L}_{HNN}^{CP} =g^{CP}_{HNN} \, H \, \bar{N}N\,, 
\end{align}
with the ansatz that the coupling is made via pion loops as shown in Fig.~\ref{fetaNloop}.

The chiral Lagrangians to describe the couplings appearing in those loops are
\begin{align}
\nn\mathcal{L}_N^{(1)}=&
\bar{\Psi}\left(\mathrm{i}\sla{\mathrm{D}}-m+\frac{g_0}{2}\sla u\gamma_5\right)\Psi,\\
\mathcal{L}^{(1)}_{\Delta\pi N}=&\frac{\mathrm{i}h_A}{2F_0M_\Delta}\bar{\Psi}T^a
\gamma^{\mu\nu\lambda}(\partial_\mu\Delta_\nu)(\partial_\lambda\pi^a) + \text{h.c.},\label{ELag}
\end{align}
where $\Psi$ is the nucleon doublet $\left(p,n\right)^T$ with mass $m$, and $\Delta$ 
the isospin-3/2 quadruplet $\left(\Delta^{++},\Delta^{+},\Delta^{0},\Delta^{-}\right)^T$ 
with mass $M_\Delta$. The covariant derivative is given by
\begin{align}
\mathrm{D}_\mu=\left(\partial_\mu+\Gamma_\mu\right),\quad \Gamma_\mu=
\frac12\left[u^\dagger(\partial_\mu-\mathrm{i}r_\mu)u
+u(\partial_\mu-\mathrm{i}l_\mu)u^\dagger\right]. 
\end{align}
The meson fields appear through
\begin{align}
u^2=&U=\exp\left(\frac{\mathrm{i}\Pi}{F_0}\right), \quad
 \Pi=\left(\begin{array}{cc}
\pi^0&\sqrt2\pi^+\\
\sqrt2\pi^-&-\pi^0\label{pimatrix}
\end{array}\right),\quad
u_\mu=\mathrm{i}\left[u^\dagger(\partial_\mu-\mathrm{i}r_\mu)u
-u(\partial_\mu-\mathrm{i}l_\mu)u^\dagger\right],
\end{align}
with $r_\mu$ and $l_\mu$ being right- and left-handed external fields, and 
$F_0$ is the meson-decay constant.
At leading chiral order, the \gls{lec} $g_0$ corresponds to the physical axial-vector 
coupling constant $g_A=1.27$. Furthermore, we use the notation
\begin{align}
\gamma^{\mu\nu\lambda}=\frac12\left\{\gamma^{\mu\nu},\gamma^\lambda\right\},\quad \gamma^{\mu\nu}=\frac12\left[\gamma^\mu,\gamma^\nu\right].
\end{align}
The coupling $h_A$ can be obtained from the $\Delta$ width, 
leading to the value $h_A=2.85$~\cite{Pascalutsa:2004je}. 
The conventions and definitions for the isospin operators $T^i$ follow 
Ref.~\cite{Pascalutsa:2006up}:
\begin{align}
\nn T^1=&\frac{1}{\sqrt 6}\left(
\begin{array}{cccc}
-\sqrt 3 &0&1&0
\\
0&-1&0&\sqrt 3
\end{array}
\right)\,,\\
\nn T^2=&\frac{-\mathrm{i}}{\sqrt 6}\left(
\begin{array}{cccc}
\sqrt 3 &0&1&0
\\
0&1&0&\sqrt 3
\end{array}
\right)\,,\\
T^3=&\sqrt{\frac23 }\left(
\begin{array}{cccc}
0 &1&0&0
\\
0&0&1&0
\end{array}
\right)\,.
\end{align}

From isospin considerations, it should be clear that there is no contribution 
from Fig.~\ref{fetaNloopb}, due to the cancellation of the $\pi^+$ and the $\pi^-$ loop. 
The loops in Figs.~\ref{fetaNloopa} and \ref{fetaNloopaD} do contribute, though. With the Lagrangians 
introduced in Eq.~\ref{ELag} and the vertex from Eq.~\ref{ELagCPepp}, 
the loop in Fig.~\ref{fetaNloopa} reads
\begin{align}
g^{CP}_{H NN}=-\mathrm{i}I_{N\pi}^2\frac{g_A^2f_{H\pi\pi}M_H}{F_\pi^2}
\int{\frac{\mathrm{d}^dz}{(2\pi)^d}}
\frac{(\slz+\slk)\gamma_5(\slp-\slz+m)\slz\gamma_5}
{[(k+z)^2-M_\pi^2][z^2-M_\pi^2][(p-z)^2-m^2]}\,,
\end{align}
where $k$ is the momentum of the $\eta(\eta')$, and $I_{N\pi}$ is the isospin factor. 
It is $1/2$ for the $\pi^0n$ loop and 
$\sqrt{2}/2$ for $\pi^-p$. The incoming nucleon momentum is given by $p$, $M_\pi$ 
is the pion mass, and $m$ the nucleon mass. 
To estimate the coupling, we use the approximation where the external nucleon legs 
are on-shell. When simplifying this integral with the help of Feynman parameters and 
dimensional regularization, we obtain the result: 
\begin{align}
\nn g^{CP}_{H NN}=&\frac{g_A^2I_{N\pi}^2}{F_\pi^2}f_{H\pi\pi}
M_H\int_0^1\mathrm{d}f_a\int_0^{1-f_a}\mathrm{d}f_b
\Bigg[-3m(f_b +1)\lambda_2(\Delta_{\eta NN})\\[1em]
&+m\frac{2+f_b}{2}\rho_2(\Delta_{\eta NN})+\Bigg(f_a(f_b +2)(f_a+f_b-1) k^2 m+f_b^3 m^3\Bigg)\lambda_3(\Delta_{\eta NN})\Bigg]\,.
\label{EAnalyticEtaNN}
\end{align} 
In the last expression, $f_a$ and $f_b$ are Feynman parameters, and
\begin{align}\label{EqDimRegFuncs}
\nn\lambda_2(\Delta)&=\frac{1}{16\pi^2}
\left[\frac{2}{\epsilon}-\log\left(\frac{\Delta}{\mu^2}\right)
+\log(4\pi)-\gamma_E\right]\,,\\[1em]
\nn\rho_2(\Delta)&=\frac{2}{16\pi^2}\,,\\[1em]
\lambda_3(\Delta)&=\frac{1}{16\pi^2\Delta}\,,
\end{align}
with $\epsilon=4-D$. Here, $D$ is the Minkowski-space dimension, 
and the renormalization scale $\mu$ is set to the nucleon mass. 
For this diagram, we have 
\begin{align}
\Delta_{H NN}=M_\pi^2(1-f_b)-f_a k^2(1-f_a-f_b)+m^2 f_b^2\,.
\end{align} 

For the purpose of comparison, we extract the heavy-baryon limit 
from Eq.~\ref{EAnalyticEtaNN} by taking the leading order of the Taylor expansion 
around the small parameter $m^{-1}$. 
When choosing a vanishing $k^2=0$ for the $\eta$, 
the $CP$-violating coupling has the compact form
\begin{align}
g^{CP,\,HB}_{H NN}=\frac{3 f_{\eta \pi \pi} g_A^2 m M_H 
(\gamma_E-2-\log (4 \pi )-\frac 2\epsilon)}{32 \pi ^2 F_\pi^2}\,.
\end{align}
This result is in agreement with the previous calculations of Ref.~\cite{Gorchtein:2008pe} 
after some typos are corrected.

The divergences are absorbed with the $\overline{MS}$ scheme: 
terms proportional to $\frac{2}{\epsilon}+\log(4\pi)-\gamma_E$ are subtracted. 
Setting $k^2=0$, we obtain a compact 
result for Eq.~\ref{EAnalyticEtaNN}: 
\begin{align}
\nn &g^{CP}_{H NN}=-\frac{3 f_{H \pi \pi} g_A^2 M_H}{16 \pi ^2 F_\pi^2 m } 
\Bigg\{M_\pi^2 \log \left(\frac{m}{M_\pi}\right)+m^2
+\frac{M_\pi \left(M_\pi^2-3 m^2\right)}{\sqrt{4 m^2-M_\pi^2}}\\[1em]
&\times\Bigg[ \arctan\left(\frac{M_\pi}{\sqrt{4 m^2-M_\pi^2}}\right)
+\arctan\left(\frac{2 m^2-M_\pi^2}{\sqrt{4 m^2 M_\pi^2-M_\pi^4}}\right)\Bigg]\Bigg\}\,.
\end{align}

As for Fig.~\ref{fetaNloopaD}, with a $\Delta$ intermediate state, the coupling reads
\begin{align}
\nn g^{CP}_{H NN,\,\Delta}=&\mathrm{i}I_{N\Delta\pi}^2
\frac{h_A^2f_{H\pi\pi}M_H}{F_\pi^2M_\Delta^2}
\int{\frac{\mathrm{d}^dz}{(2\pi)^d}}\\[1em]
&\times
\frac{(p^\alpha-z^\alpha)z^\delta\gamma_{\alpha\beta\delta}
S_\Delta^{\beta\beta'}(p-z)(p^{\alpha'}-z^{\alpha'})
(z^{\delta'}+k^{\delta'})\gamma_{\alpha'\beta'\delta'}}
{[(k+z)^2-M_\pi^2][z^2-M_\pi^2][(p-z)^2-M_\Delta^2]}\,,
\end{align}
where the isospin factor $I_{N\Delta\pi}$ is $1/6$ for the $\pi^0\Delta^0$ loop 
and $1/3$ for the combination of $\pi^-\Delta^+$ and $\pi^+\Delta^-$. 
The $\Delta$ propagator is 
\begin{align}
\nn S_\Delta^{\alpha\beta}(p)=&
\frac{\slp+M_\Delta}{p^2-M_\Delta^2+\mathrm{i}\varepsilon}
\bigg[
-g^{\alpha\beta}
+ \frac{1}{D-1}\gamma^\alpha\gamma^\beta\\[1em]
&+ \frac{1}{(D-1)M_\Delta}(\gamma^\alpha p^\beta - \gamma^\beta p^\alpha)
+ \frac{D-2}{(D-1)M_\Delta^2}p^\alpha p^\beta
\bigg]\,. 
\label{EqDelPropDef}
\end{align} 
When putting the external nucleons on shell, and choosing $k^2=0$, we obtain: 
\begin{align}
\nn &g^{CP}_{H NN,\,\Delta}=\frac{f_{H \pi \pi}
h_A^2 m^2 M_H}{1152 \pi ^2 F_\pi^2 M_\Delta^2}
\Bigg\{-\frac{6 \left(m^2+M_\pi^2-M_\Delta^2\right)}{m}
+6 (2 m+3 M_\Delta) \log
   \left(\frac{M_\Delta^2}{m^2}\right)+4 m\\[1em]
\nn&-\frac{6 \left(2 m^4+3 m^3 M_\Delta+m^2 
\left(2 M_\Delta^2-6 M_\pi^2\right)+m \left(3 M_\Delta^3-3 M_\pi^2 M_\Delta\right)+2
   \left(M_\pi^2-M_\Delta^2\right)^2\right)}{m^3}\\[1em]
\nn&+\frac{1}{m^5 \sqrt{-m^4+2 m^2 \left(M_\pi^2+M_\Delta^2\right)
-\left(M_\pi^2-M_\Delta^2\right)^2}}\\[1em]
\nn&\times\Bigg[6 \left(2 m^4-5 m^3 M_\Delta+m^2 
\left(6 M_\Delta^2-4 M_\pi^2\right)+5 m M_\Delta \left(M_\pi^2-M_\Delta^2\right)+2
   \left(M_\pi^2-M_\Delta^2\right)^2\right)\\[1em]
\nn& \times\left(m^2+2 m M_\Delta-M_\pi^2+M_\Delta^2\right)^2\Bigg]\\[1em]
\nn&\times\Bigg[ \arctan\left(\frac{-m^2-M_\pi^2+M_\Delta^2}{\sqrt{-m^4+2 m^2
   \left(M_\pi^2+M_\Delta^2\right)-\left(M_\pi^2-M_\Delta^2\right)^2}}\right)\\[1em]
\nn&- \arctan\left(\frac{m^2-M_\pi^2+M_\Delta^2}{\sqrt{-m^4+2 m^2 
\left(M_\pi^2+M_\Delta^2\right)-\left(M_\pi^2-M_\Delta^2\right)^2}}\right)\Bigg]\\[1em]
\nn&+\frac{3 \log \left(\frac{M_\pi^2}{M_\Delta^2}\right)}{m^5}
\Bigg[2 m^6+3 m^5 M_\Delta+6 m^4 M_\pi^2+6 m^3 M_\pi^2
   M_\Delta+6 m^2 M_\pi^2 \left(M_\Delta^2-M_\pi^2\right)\\[1em]
&-3 m M_\Delta \left(M_\pi^2-M_\Delta^2\right)^2+2 \left(M_\pi^2-M_\Delta^2\right)^3\Bigg]\Bigg\}\,.
\end{align}

Using the pion-decay ratio $F_\pi=92.4~\unit{MeV}$, 
and the upper bounds on 
the $CP$-violating couplings $f_{H\pi\pi}$ 
as introduced in Eq.~\ref{upperbound}, one obtains 
the following upper limits: 
\begin{align}
\nn &|g^{CP}_{\eta NN}|=2.8\cdot 10^{-5},~|g^{CP}_{\eta' NN}|=5.1\cdot 10^{-4},\\[1em]
\nn &|g^{CP,\,HB}_{\eta NN}|=3.9\cdot 10^{-5},~|g^{CP,\,HB}_{\eta' NN}|=7.1\cdot 10^{-4},\\[1em]
&|g^{CP}_{\eta NN,\,\Delta}|=7.9\cdot 10^{-6},~|g^{CP}_{\eta' NN,\,\Delta}|=1.4\cdot 10^{-4}\,.
\end{align}
One can see from the numerical result that it is important to take the $\Delta$ loop 
into account, as its contribution is larger than $20\%$ of the nucleon's. 
Furthermore, although the magnitude of the heavy-baryon calculation is 
similar in size to the fully covariant one, one can see that there is 
a sizeable change of around $30\%$ in the numerical value due to this 
non-relativistic approximation.{\footnote{These couplings, without the $\Delta$ contribution, 
had been calculated previously in the \gls{hbchpt} approach, 
in Ref.~\cite{Gorchtein:2008pe}. A direct comparison of the numerical results has 
little meaning because of some errors in the formulas. Also, now we have experimentally better 
constrained values for the branching ratios of Eq.~\ref{Epdgbranch}~\cite{Agashe:2014kda,Aaij:2016jaa}.}

\section{The $CP$-conserving coupling of the $\eta$ and 
$\eta'$ to the nucleon}\label{STheoetaNN}

The $CP$-conserving coupling of the $\eta(\eta')$ to the nucleon is given by
\begin{align}
\mathcal{L}_{H NN}=-\mathrm{i}\frac{g_{H NN}}{2F_\eta}\, 
H \, \bar{N}\slk\gamma_5 N\,,
\end{align} 
with $k$ the $\eta$ momentum, and $\bar{N}$ and $N$ the outgoing and incoming 
nucleon states, respectively. In this calculation, we set the decay constant 
$F_\eta$ to the physical $SU(3)$ average of $108~\unit{MeV}$~\cite{Ledwig:2014rfa}.

The physical $\eta$ and $\eta'$ 
are a mixing of the singlet and the octet states. Thus, the coupling $\pi^0 nn$ 
is given by $-g_A=-(F+D)$, while the $\eta nn$ and $\eta' nn$ vertices have the couplings 
$g_{\eta nn} =  (D+F) \cos\psi+\sqrt{2}(F-D) \sin \psi$ and $g_{\eta' nn}= \sqrt{2}(D-F) \cos\psi+  (F+D)\sin\psi$, respectively. 
The mixing angle $\psi$ between the $\eta$ and the $\eta'$ has been estimated 
in many works~\cite{Feldmann:1998vh,Goity:2002nn,Aubert:2006cy,Ambrosino:2009sc,%
Mathieu:2009sg,Escribano:2015nra,Osipov:2015lva} to be in a range between 
$38^o$~\cite{Escribano:2015nra} from $\eta\rightarrow e^+e^-\gamma$ decay data 
and $45^o$~\cite{Goity:2002nn} in a \gls{chpt} analysis. 
The more recent results tend to have values close to $40^o$, which we use in the following. 
We also take the physical-average values for $F=0.47$ and $D=0.8$~\cite{Ledwig:2014rfa}.

\section{Couplings of vector mesons}
\label{STheoVec}

In the present work, we also study the effects of loops containing vector mesons 
coupling to the $\eta(\eta')$ and to the nucleon. 
The relevant pieces of the Lagrangians describing this type of 
couplings are~\cite{Drechsel:1998hk,Chiang:2001as} 
\begin{align}
\mathcal{L}_{\gamma H V}=&-\frac{e\lambda_V}{4M_H} \, 
\epsilon_{\mu\nu\alpha\beta} \, F^{\mu\nu} \, V^{\alpha\beta}\eta,\\[1em]
\mathcal{L}_{VNN}=&\bar{N} \, 
\left(g_v\gamma^\mu+g_t\frac{\sigma^{\mu\nu}}{2m}\partial_\nu\right) 
\, V_\mu \, \tau_V \, N,
\end{align}
where the values taken for the coupling constants~\cite{Gorchtein:2008pe,Chiang:2001as} 
are summarized in Table~\ref{tvecval}. 
The electromagnetic field couples via the usual definition 
$F^{\mu\nu}=\partial^{\mu}\mathcal{A}^{\nu}-\partial^{\nu}\mathcal{A}^{\mu}$, 
and 
$V^{\mu\nu}=\partial^\mu V^\nu - \partial^\nu V^\mu$. 
The propagator of a vector-meson field with momentum $k$ and mass $m_V$ is taken as
\begin{align}
\frac{1}{k^2-m_V^2}\left(-g^{\alpha\beta}+\frac{k^\alpha k^\beta}{m_V^2}\right).
\end{align}
\begin{table}[htbp]
\begin{center}
\begin{tabular}{c|cccccc}
$V$& $g_v^V$ & $g_t^V/g_v^V$ & $\lambda_V$ 
& $\lambda'_V$ &$\tau_V$\\
\hline
$\rho^0$&$2.4$&$6.1$&$0.9$&$1.18$&$\tau_3$\\
$\omega$&$16$&$0$&$0.25$&$0.43$&$1$
\end{tabular}
\caption{Parameters for the vector-meson coupling Lagrangians. $\tau_3$ assumes the values $1$ and $-1$ for the proton and the neutron, respectively.}
\label{tvecval}
\end{center}
\end{table} 

Here we want to remark that the values for the couplings are poorly known, 
for which reason they are an important source of uncertainty for the results. 
Furthermore, in higher orders they have a dependency on the virtuality of 
the vector meson, which we ignore in the leading-order calculations that follow.

\section{Calculation of the nucleon EDM}
\label{STheoEDM}

The \gls{edm} is extracted from the amplitude coupling the photon to the nucleon. 
In our case, as the amplitude always involves a $CP$-violating vertex, only one form factor containing the EDM appears. 
Therefore, the $CP$-violating part of the vector current $J^\mu$ between baryon states reads: 
\begin{align}
\bra{B(p')}J^\mu_\text{CPV}\ket{B(p)} 
=\bar{u}(p')\frac{\mathrm{i}\sigma^{\mu\nu}\gamma_5q_\nu}{4m} F_\text{EDM}(q^2)u(p)\,,
\end{align} 
where $q_\nu$ is the photon momentum, $\epsilon_\mu$ its polarization, and 
$\sigma^{\mu\nu}=\mathrm{i}\,\gamma^{\mu\nu}$. 
At the point where $q^2=0$, 
the form factor reduces to the electric dipole moment 
$F_\text{EDM}(0) = \tilde{d}_N$. 
In our model, the CPV comes from the loops of Fig.~\ref{fedm}. 
\begin{figure}[htbp]
\begin{center}
\subfigure[]{
\label{fedma}
\includegraphics[width=0.3\textwidth]{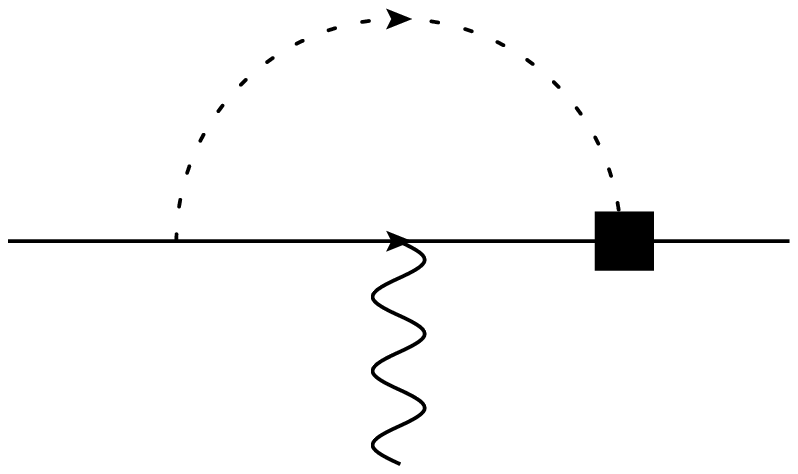}\hspace{2cm}
\includegraphics[width=0.3\textwidth]{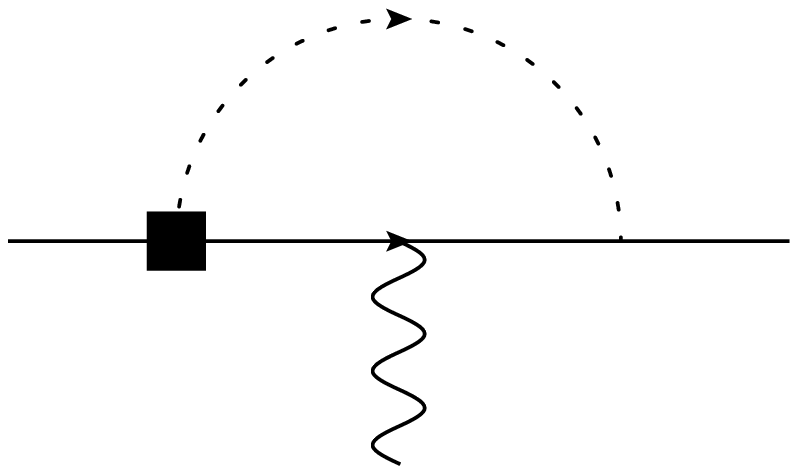}}\\
\subfigure[]{
\label{fedmb}
\includegraphics[width=0.3\textwidth]{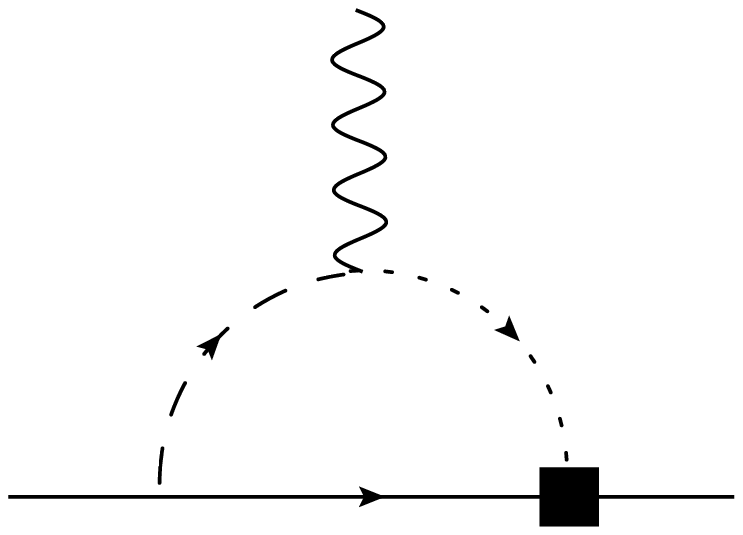}\hspace{2cm}
\includegraphics[width=0.3\textwidth]{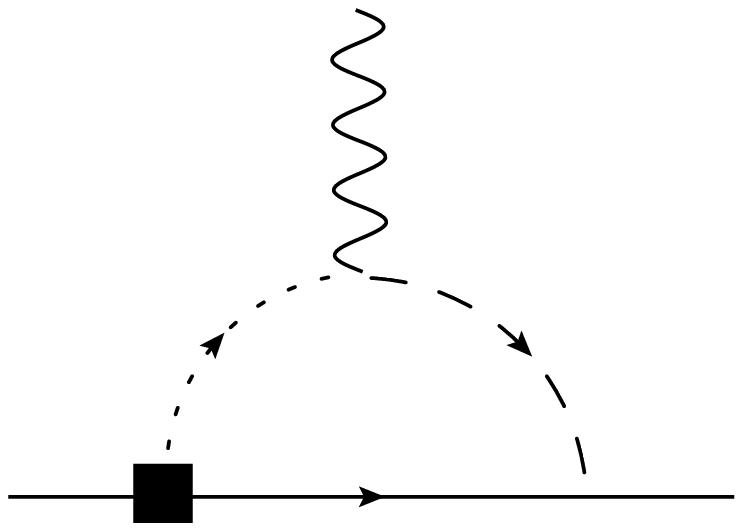}}
\end{center}
\caption{Loops that can contribute to the neutron EDM. 
The solid line represents the neutron, the dotted line the $\eta(\eta')$, 
the dashed lines are vector-meson contributions, and the wavy line corresponds 
to the photon. Again, the black box stands for a $CP$-violating vertex.}
\label{fedm}
\end{figure}

In Fig.~\ref{fedma}, the photon couples to the nucleon that propagates inside the loop. 
In the particular case of the neutron, 
the leading-order coupling to the photon vanishes, for which reason only 
next-to-leading order terms contribute. The second-order nucleon Lagrangian is 
needed to describe such a vertex at lowest non-vanishing order for the neutron, 
which reduces to
\begin{align}
\mathcal{L}^{(2)}_{\gamma nn}=\sigma^{\mu\nu}F_{\mu\nu}
\frac{e\kappa_n}{4m}\,,
\end{align} 
where $\kappa_n=-1.913$ 
is the neutron magnetic moment in units of $\frac{e}{2m}$.

A direct coupling of the photon to an $\eta(\eta')$ propagating inside 
a loop is not possible, due to this meson's vanishing charge. 
Nevertheless, as is depicted in Fig.~\ref{fedmb}, it is possible to achieve 
a coupling via a vector-meson exchange, which here we also perform for 
the sake of comparison with Ref.~\cite{Gorchtein:2008pe}, 
and for an estimate of the importance of its effect.

The amplitude of Fig.~\ref{fedma} reads 
\begin{align}
\nn\frac{e\, \kappa_n\, g_{H NN}\,\bar{g}_{H NN}}{8m F_\eta}
\int{\frac{\mathrm{d}^dz}{(2\pi)^d}}\frac{1}{[z^2-M_H^2]
[(p-z)^2-m^2][(p+q-z)^2-m^2]}\\[1em]
\nn\times\left[(\slp+ \slq-\slz+m)(\slq\sle-\sle\slq)(\slp-\slz+m)
\slz\gamma_5\right.\\[1em]
\left.-\slz\gamma_5(\slp+ \slq-\slz+m)
(\slq\sle-\sle\slq)(\slp-\slz+m)\right]\,,
\end{align} 
which for the dipole moment $\tilde{d}_n$ 
in units of $\frac{e}{2m}$ at $q^2=0$ leads to 
\begin{align}
\nn \tilde{d}_{n,a}=&\frac{ m\, \kappa_n\,\bar{g}_{H NN}\,{g}_{H NN}}{F_H} \, 
\int_{0}^1\mathrm{d}f_b\int_{0}^{1-f_b}\mathrm{d}f_a
\Bigg\{(6f_a-5)\lambda_2(\Delta_\text{EDM,a})+(2-f_a)\rho_2(\Delta_\text{EDM,a})\\[1em]
&-2m^2\left[f_a(f_a^2-2)+2(1+(f_a-1)f_a)f_b+(f_a-2)f_b^2\right]\lambda_3(\Delta_\text{EDM,a})
\Bigg\} .
\end{align}
Together with the definitions in Eq.~\ref{EqDimRegFuncs}, we choose the notation
\begin{align}
\Delta_\text{EDM,a}=m_\eta^2(1-f_a-f_b)+m^2(f_a+f_b^2).
\end{align} 
After integration, the analytical expression is also quite compact: 
\begin{align}
\nn\tilde{d}_{n,a}=&\frac{ \kappa_n\,\bar{g}_{H NN}\, g_{H NN}}{32 \pi ^2 
\text{F$\eta $} m^3} \Bigg\{ m^4-3 m^2 m_\eta ^2+
\left(3 M_H^4-6 m^2 M_H^2\right) \log
   \left(\frac{M_\eta }{m}\right)\\[1em]
\nn&-\frac{3 m_H^3}{ \sqrt{4 m^2-M_H ^2}} \left(M_H^2-4 m^2\right)\\[1em]
&\times \left[\arctan\left(\frac{M_H}{\sqrt{4m^2-M_H^2}}\right)
+ \arctan\left(\frac{2 m^2-M_H^2}{M_H \sqrt{4m^2-M_H^2}}\right)\right]\Bigg\}.
\end{align} 

As for the amplitude in F.~\ref{fedmb}, it is given by 
\begin{align}
\nn\frac{e\,\lambda_V\,\tau_V \,\bar{g}_{H NN}}{m_H}
\int{\frac{\mathrm{d}^dz}{(2\pi)^d}} 
\Bigg[\frac{(\slz+m)}{[(p-z)^2-M_V^2][z^2-m^2][(p'-z)^2-M_H^2]}\\[1em]
\nn\times q_\mu\epsilon_\nu (p-z)_\alpha\mathrm{i}\epsilon^{\mu\nu\alpha\beta}
\left(-g_{\beta\beta'}+\frac{(p-z)_\beta (p-z)_{\beta'}}{M_V^2}\right)
\left(g_v^V\gamma^{\beta'} - \frac{g_t^V}{4m} (p-z)^{\alpha'}
[\gamma^{\beta'},\gamma^{\alpha'}]\right)\\[1em] 
\nn-\frac{
\left(g_v^V\gamma^{\beta'} + \frac{g_t^V}{4m} (p'-z)^{\alpha'}
[\gamma^{\beta'},\gamma^{\alpha'}]\right)\left(-g_{\beta'\beta}
+\frac{(p'-z)_{\beta'} (p'-z)_{\beta}}{M_V^2}\right)}{[(p-z)^2-M_H^2]
[z^2-m^2][(p'-z)^2-M_V^2]}\\[1em]
\times q_\mu\epsilon_\nu (p'-z)_\alpha\mathrm{i}\epsilon^{\mu\nu\alpha\beta}
(\slz+m)\Bigg]\,.
\end{align}
For this loop diagram, the analytical result has the very simple form
\begin{align}
\nn \tilde{d}_{n,b}=&2\frac{\lambda_V \,\tau_V\, m\,\bar{g}_{\text{$\eta $NN}} 
}{M_H} \int_{0}^1\mathrm{d}f_b\int_{0}^{1-f_b}\mathrm{d}f_a
\Bigg\{(g_v^V-g_t^V)[2\lambda_2(\Delta_\text{EDM,b})+3\rho_2(\Delta_\text{EDM,b})]
\Bigg\}\,,
\end{align} 
where
\begin{align}
\nn\Delta_\text{EDM,b}&=m^2(1-f_a-f_b)^2+M_H^2f_b+m_V^2f_a.
\end{align}
Note that, for each of the two diagrams in F.~\ref{fedmb} separately, 
there are also pieces of the type $\lambda_3(\Delta_\text{EDM,b})$, 
but they cancel each other. Integrating over the Feynman parameters yields 
\begin{align}
\nn\tilde{d}_{n,b}=&\frac{\bar{g}_{H NN}(g_t^V-g_v^V)\lambda_V\tau_V}{24 \pi ^2 
m^3 M_H(M_H^2-m_V^2)} \Bigg\{
m^2(m_\eta^4-m_V^4)\\[1em]
\nn&+M_H^3(4m^2-M_H^2)^{3/2}\Bigg[\arctan
\left(\frac{M_H}{\sqrt{4m^2-M_H^2}}\right)
-\arctan\left(\frac{M_H^2-2m^2}{M_H\sqrt{4m^2-M_H^2}}\right)\Bigg]\\[1em]
\nn&-M_V^3(4m^2-M_V^2)^{3/2}\Bigg[\arctan\left(\frac{M_V}
{\sqrt{4m^2-M_V^2}}\right)
-\arctan\left(\frac{M_V^2-2m^2}{M_V\sqrt{4m^2-M_V^2}}\right)\Bigg]\\[1em]
&+M_H^4(6m^2-M_H^2)\log\left(\frac{M_H}{m}\right)-M_V^4(6m^2-M_V^2)
\log\left(\frac{M_V}{m}\right)
 \Bigg\}\,.
\end{align}

The numerical results are summarized 
in Table~\ref{TresEDM}. The vector-meson contributions of Fig.~\ref{fedmb} turn out 
to be of the same order of magnitude as the loops in Fig.~\ref{fedma}. 
This is to be expected, even though the vector mesons are higher-mass states. 
For Fig.~\ref{fedma}, the Lagrangian of first chiral order does not allow a coupling 
of the photon to the neutron. Therefore, this contribution is suppressed, and 
the vector-meson contributions become equally important. The sum of all the 
contributions yields a total value for the dipole moment of 
$d_n^\text{tot}=4.3\cdot 10^{-18}e~\unit{cm}$. Note that this value takes into account 
the new result for the $\eta'$ two-pion 
decay~\cite{Aaij:2016jaa}. Therefore it is smaller by approximately a factor $\sqrt{3}$, 
when compared to values obtained from the $\eta'$ two-pion 
decays in Ref.~\cite{Agashe:2014kda}. 
Considering the current experimental upper limit of $2.9\cdot 10^{-26}e~\unit{cm}$ 
for the neutron dipole moment, the ratio between expectation and measurement 
is of the order of $10^{8}$. This means that the present upper limit 
for the decay ratio of the $\eta(')$ into two pions gives a large overestimation 
of the $CP$-violating coupling constant. In fact, in order for the results to be 
compatible with the experimental constraint on the neutron dipole moment, the 
branching ratio would have to be at least eight orders of magnitude smaller. 
\begin{table}[htbp]
\begin{center}
\begin{tabular}{c|cc}
&$\eta$&$\eta'$\\\hline
Fig.~\ref{fedma}&$3.1\cdot 10^{-20}$&$1.5\cdot 10^{-18}$\\
Fig.~\ref{fedmb}&$2.1\cdot 10^{-19}$&$2.6\cdot 10^{-18}$
\end{tabular}
\end{center}
\caption{Contributions to the upper limit of the neutron \gls{edm}, 
from the current experimental upper limits of the $\eta$ and $\eta'$ 
branching ratios into two pions~\cite{Agashe:2014kda,Aaij:2016jaa}. 
The units are $e~\unit{cm}$.}
\label{TresEDM}
\end{table}

It is interesting to confront these results with those in Ref.~\cite{Pich:1991fq}. 
There, as mentioned, the size of the neutron \gls{edm} was estimated within 
a similar framework as presented here, but by considering a $CP$-violating vertex in the coupling 
of the charged mesons to the baryons, and calculating their induced contributions to the EDM at leading chiral order. 
There, up to a factor including the unknown $CP$-violating phase 
$\theta$, the \gls{edm} was estimated to be of the order of 
$10^{-16}e~\unit{cm}$. The fact that we get an estimate 
approximately two orders of magnitude smaller is in good agreement with that calculation, 
knowing that for the neutral mesons considered here the diagrams that contribute are of 
the next chiral order.

It is important to keep in mind that the values shown in Table~\ref{TresEDM} are not to be 
seen as predictions for the neutron \gls{edm}, but as estimates for the order of magnitude 
of the $\eta(\eta')$ branching ratios into two pions. Other processes, which are beyond the scope of 
this paper, give additional contributions to the neutron \gls{edm}. 
These are, e.g., pieces obtained from the $CP$-violating decay of the $\eta'$ into 
four pions, or processes that do not conserve flavour via the quark-mixing matrix. 
Furthermore, as mentioned in Sec.~\ref{STheoVec}, some of the coupling constants used here 
are poorly known, and the results depend on the renormalization scheme used. 
Nevertheless, due to the very large discrepancy between the experimental constraint on 
the \gls{edm} and the one calculated from the current upper limits for the $CP$-violating 
branching ratios, the results are still rigorous enough to be instructive. 
The conclusions made here remain, even if other processes are to be additionally 
considered, or if the coupling constants are to have different sizes.

\section{Summary}
\label{Ssummedm}

In the present paper, we calculated the nucleon \gls{edm} originated by a $CP$-violating 
coupling to the $\eta(\eta')$ meson. In particular, we focused on the result for the 
neutron, as its experimental upper limit is very small, $2.9\cdot 10^{-26}e~\unit{cm}$. 
This limit sets a very strong constraint on observables related to it. More specifically, 
if a neutron \gls{edm} is to exist, then $CP$ violation has to occur. 
Therefore, here the goal was to give an estimate of the size of this violation.

This was achieved by constructing a $CP$-violating coupling of the $\eta$ to the nucleon 
via loops that include an $\eta(\eta')\pi\pi$ vertex. While there are experimental 
results for the upper limit of the $\eta(\eta')\rightarrow\pi\pi$ decay ratio, 
here we wanted to test if this constraint is indeed compatible with the limit on 
the neutron \gls{edm}. The $\Delta$-isobar
contributions were taken into account as well, leading to a correction to 
the $CP$-violating $\eta(\eta') NN$ vertex larger than $20\%$.

We considered two possible sources for the neutron \gls{edm}. 
In one case, the photon coupled to the neutron within a loop with a $CP$-violating 
$\eta NN$ vertex. In the other, vector-meson contributions were 
considered as well. The two contributions turned out to be of a similar size.

In total, we obtained a constraint on the $CP$-violating $\eta(\eta')\rightarrow\pi\pi$ 
decay ratio roughly eight orders of magnitude smaller than measured in the experiment 
so far. This is a very instructive result, since it gives an estimate on symmetry 
violations in nature, where experimental results are not yet achievable.

\newpage
\section*{Acknowledgements}

This research has been partially supported by the Spanish 
Ministerio de Econom\'{\i}a y Competitividad (MINECO) 
and the European fund for regional development (FEDER) 
under Contracts FIS2014-51948-C2-2-P and SEV-2014-0398, and by Generalitat 
Valenciana under Contract PROMETEOII/2014/0068. 
A.N.H.B. acknowledges support from the Santiago Grisol\'{\i}a 
program of the Generalitat Valenciana, and thanks Michael Gorchtein for valuable discussions. 
This work was supported
by the German Bundesministerium f\"ur Bildung und Forschung (BMBF)
under Project 05P2015 - ALICE at High Rate (BMBF-FSP 202):
``Jet- and fragmentation processes at ALICE and the parton structure  
of nuclei and structure of heavy hadrons'', by CONICYT (Chile)  Basal No. FB082 and Ring No.~ACT1406,
by FONDECYT  (Chile) Grants No.~1140471 and No.~1150792,
by Tomsk State University Competitiveness
Improvement Program and the Russian Federation program ``Nauka''
(Contract No. 0.1526.2015, 3854). Also, A.S. is grateful for the fact that his work was supportedd by the Russian Science Foundation Grant (RSCF 15-12-10009).


\begin{thebibliography}{99}

\bibitem{Agashe:2014kda} 
K.~A.~Olive et al. (Particle Data Group), \emph{Review of Particle Physics},
Chin.\ Phys.\ C \textbf{38}, 090001 (2014).

\bibitem{Pich:1991fq} 
A.~Pich and E.~de Rafael, \emph{Strong CP violation in an effective chiral Lagrangian approach},
Nucl.\ Phys.\ B \textbf{367}, 313 (1991).

\bibitem{Dar:2000tn} 
S.~Dar, \emph{The Neutron EDM in the SM: A Review},
hep-ph/0803.2906 (2008).

\bibitem{Gorchtein:2008pe} 
M.~Gorchtein, \emph{Nucleon EDM and rare decays of $\eta$ and $\eta^\prime$ mesons},
hep-ph/0008248 (2000).

\bibitem{Jarlskog:1995ww} 
C.~Jarlskog and E.~Shabalin, \emph{How large are the rates of the CP violating $\eta$, $\eta^\prime \rightarrow \pi\pi$ decays?},
Phys.\ Rev.\ D \textbf{52}, 248 (1995).

\bibitem{Peccei:1977hh} 
R.~D.~Peccei and H.~R.~Quinn, \emph{CP Conservation in the Presence of Instantons},
Phys.\ Rev.\ Lett.\ \textbf{38}, 1440 (1977).

\bibitem{Peccei:1977ur} 
R.~D.~Peccei and H.~R.~Quinn, \emph{Constraints Imposed by CP Conservation in the Presence of Instantons},
Phys.\ Rev.\ D \textbf{16}, 1791 (1977).

\bibitem{Faessler:2006at} 
A.~Faessler, T.~Gutsche, S.~Kovalenko, and V.~E.~Lyubovitskij, \emph{Implications of $R$-parity violating supersymmetry for atomic and hadronic EDMs},
Phys.\ Rev.\ D \textbf{74}, 074013 (2006).

\bibitem{Faessler:2006vi} 
A.~Faessler, T.~Gutsche, S.~Kovalenko, and V.~E.~Lyubovitskij, \emph{Hadronic electric dipole moments in R-parity violating supersymmetry},
Phys.\ Rev.\ D \textbf{73}, 114023 (2006).

\bibitem{Kuckei:2005pg} 
J.~Kuckei, C.~Dib, A.~Faessler, T.~Gutsche, S.~Kovalenko, V.~E.~Lyubovitskij, and K.~Pumsa-ard, \emph{Strong CP violation and the neutron electric dipole form-factor},
Phys.\ Atom.\ Nucl.\ \textbf{70}, 349 (2007).

\bibitem{Dib:2006hk} 
C.~Dib, A.~Faessler, T.~Gutsche, S.~Kovalenko, J.~Kuckei, V.~E.~Lyubovitskij, and K.~Pumsa-ard, \emph{The Neutron electric dipole form-factor in the perturbative chiral quark model},
J.\ Phys.\ G \textbf{32}, 547 (2006).

\bibitem{Pospelov:2005pr} 
M.~Pospelov and A.~Ritz, \emph{Electric dipole moments as probes of new physics},
Annals Phys.\ \textbf{318}, 119 (2005).

\bibitem{Aaij:2016jaa} 
R.~Aaij et al. (LHCb Collaboration), \emph{Search for the $C\!P$-violating strong decays $\eta \to \pi^+\pi^-$ and $\eta^\prime(958) \to \pi^+\pi^-$},
hep-ex/1610.0366 (2016).

\bibitem{Ghoul:2015ifw} 
H.~Al Ghoul et al. (GlueX Collaboration), \emph{First Results from The GlueX Experiment},
AIP Conf.\ Proc.\ \textbf{1735}, 020001 (2016).

\bibitem{Gegelia:1999gf} 
J.~Gegelia and G.~Japaridze, \emph{Matching heavy particle approach to relativistic theory},
Phys.\ Rev.\ D \textbf{60}, 114038 (1999).

\bibitem{Fuchs:2003qc} 
T.~Fuchs, J.~Gegelia, G.~Japaridze, and S.~Scherer, \emph{Renormalization of relativistic baryon chiral perturbation theory and power counting},
Phys.\ Rev.\ D \textbf{68}, 056005 (2003).

\bibitem{Pascalutsa:2004je} 
V.~Pascalutsa and M.~Vanderhaeghen, \emph{Magnetic moment of the $\Delta(1232)$-resonance in chiral effective field theory},
Phys.\ Rev.\ Lett.\ \textbf{94}, 102003 (2005).

\bibitem{Pascalutsa:2006up} 
V.~Pascalutsa, M.~Vanderhaeghen, and S.-N.~Yang, \emph{Electromagnetic excitation of the $\Delta(1232)$-resonance},
Phys.\ Rept.\ \textbf{437}, 125 (2007).

\bibitem{Ledwig:2014rfa} 
T.~Ledwig, J.~Martic Camalich, L.~S.~Geng, and M.~J.~Vicente Vacas, \emph{Octet-baryon axial-vector charges and SU(3)-breaking effects in the semileptonic hyperon decays},
Phys.\ Rev.\ D \textbf{90}, 054502 (2014).

\bibitem{Feldmann:1998vh} 
T.~Feldmann, P.~Kroll, and B.~Stech, \emph{Mixing and decay constants of pseudoscalar mesons},
Phys.\ Rev.\ D \textbf{58}, 114006 (1998).

\bibitem{Goity:2002nn} 
J.~L.~Goity, A.~M.~Bernstein, and B.~R.~Holstein, \emph{The Decay $\pi^0 \to \gamma \gamma$ to next to leading order in chiral perturbation theory},
Phys.\ Rev.\ D \textbf{66}, 076014 (2002).

\bibitem{Aubert:2006cy} 
B.~Aubert et. al (BaBar Collaboration), \emph{Measurement of the eta and eta-prime transition form-factors at $q^2 = 112$ GeV$^2$},
Phys.\ Rev.\ D \textbf{74}, 012002 (2006).

\bibitem{Ambrosino:2009sc} 
F.~Ambrosino et. al (BaBar Collaboration), \emph{A Global fit to determine the pseudoscalar mixing angle and the gluonium content of the $\eta^\prime$ meson},
JHEP \textbf{7}, 105 (2009).

\bibitem{Mathieu:2009sg} 
V.~Mathieu and V.~Vento, \emph{Pseudoscalar glueball and $\eta - \eta^\prime$ mixing},
Phys.\ Rev.\ D \textbf{81}, 034004 (2010).

\bibitem{Escribano:2015nra} 
R.~Escribano, P.~Masjuan, and P.~Sanchez-Puertas, \emph{The $\eta$ transition form factor from space- and time-like experimental data},
Eur.\ Phys.\ J.\ C \textbf{75}, 414 (2015).

\bibitem{Osipov:2015lva} 
A.~A.~Osipov, B.~Hiller, and A.~H.~Blin, \emph{The $\pi^0-\eta-\eta'$ mixing in a generalized multiquark interaction scheme},
Phys.\ Rev.\ D \textbf{93}, 116005 (2016).

\bibitem{Drechsel:1998hk} 
D.~Drechsel, O.~Hanstein, S.~S.~Kamalov, and L.~Tiator, \emph{A Unitary isobar model for pion photoproduction and electroproduction on the proton up to 1 GeV},
Nucl.\ Phys.\ A \textbf{645}, 145 (1999).

\bibitem{Chiang:2001as} 
W.-T. Chiang, S.-N.~Yang, L.~Tiator, and D.~Drechsel, \emph{An Isobar model for $\eta$ photoproduction and electroproduction on the nucleon},
Nucl.\ Phys.\ A \textbf{700}, 429 (2002).

\end{thebibliography}
\end{document}